\documentclass[11pt,nofootinbib,preprintnumbers]{revtex4}

\usepackage{amssymb, amsfonts, amsmath, bm}
\usepackage{color}

\usepackage{mathrsfs}
\usepackage{enumerate}
\usepackage{amsfonts}

\usepackage[
]{hyperref}
\hypersetup{%
 setpagesize=false,
 bookmarksnumbered=true,%
 bookmarksopen=true,%
 colorlinks=true,%
 linkcolor=blue,%
 citecolor=red}

\begin{document}
\title{
{\Large
Towards rotating non-circular black holes 
\\
in string-inspired gravity
}}
\author{{\large Keisuke Nakashi}}
\affiliation{Department of Physics, Rikkyo University, Toshima, Tokyo 171-8501, Japan}
\author{{\large Masashi Kimura}}
\affiliation{Department of Physics, Rikkyo University, Toshima, Tokyo 171-8501, Japan}
\date{\today}
\preprint{RUP-20-28}

\begin{abstract}
We study stationary slowly rotating black holes, 
up to quadratic order in the spin angular momentum, 
in dynamical Chern-Simons gravity and shift symmetric Einstein scalar Gauss-Bonnet gravity, as models of string-inspired gravities.
These gravity theories modify general relativity by introducing dynamical scalar fields coupled with curvature invariants. 
We show that the linear time dependence of the scalar fields is allowed from the stationarity of the effective stress energy tensors. 
However, these time dependent scalar fields yield singular behavior of the metric functions at the black hole horizons, or they are incompatible with the stationarity of the spacetimes. 
Thus, these gravity theories admit only known solutions as regular stationary solutions. 
Our results suggest the non-existence of rotating non-circular black holes in these gravity theories.

\end{abstract}
\maketitle

\newpage
\section{Introduction}
\label{sec:intro}

In general relativity, the exterior solution of vacuum, asymptotically flat, stationary and axisymmetric black hole is given by the Kerr spacetime, which is characterized by its mass and angular momentum.
This is a consequence of the unique theorem~\cite{Israel:1967wq,Robinson:1975bv,PhysRevLett.26.331}. 
If we consider modified gravity theories, 
black holes need not be described by the Kerr spacetime (see, e.g.,~\cite{Barack:2018yly} and references therein). 
Most of the stationary and axisymmetric black hole solutions found so far satisfy the circularity condition.\footnote{
Roughly speaking, the circularity condition means that 
$t-\phi$ part and $r-\theta$ part of the metric are orthogonal, 
where $t$ and $\phi$ correspond to Killing coordinates of time translation and axisymmetry, respectively,
and $r$ and $\theta$ are the other two coordinates. See Appendix \ref{app:circularity}, for details.
}
In general relativity, it is shown that the circularity condition holds for the asymptotically flat spacetimes in vacuum system (see, e.g.,~\cite{heusler_1996,wald1984general}), the Einstein real scalar system~\cite{Herdeiro:2019oqp} and Einstein--Maxwell system~\cite{Carter:1969zz}. 
Yet, in modified gravity theories, 
it is not obvious whether spacetimes satisfy the circularity condition or not. 
If we can observe the effect of non-circularity, it directly suggests the violation of general relativity
because the circularity condition should hold for general relativity.

While the no-hair theorem holds for Einstein real scalar systems~\cite{Herdeiro:2019oqp, Bekenstein:1995un} 
and a class of modified gravity theories with stationary scalar fields~\cite{Hui:2012qt}, it is not obvious for other systems. 
For some Einstein complex scalar systems, 
hairy stationary circular black hole solutions have been found, for example,~\cite{Herdeiro:2014goa,Herdeiro:2014ima,Herdeiro:2015gia} (see also~\cite{Herdeiro:2015waa} and references therein), where scalar fields have the harmonic time dependence.\footnote{
In~\cite{Smolic:2015txa}, 
the possible time dependence of scalar fields whose 
stress energy tensors are stationary is studied.
}
In case of modified gravity theories, the linear time dependence of the scalar field has been introduced to obtain hairy static black hole solutions~\cite{Babichev:2013cya}. 
Moreover, in~\cite{VanAelst:2019kku}, it is indicated that time dependence of scalar fields is crucial to obtain non-circular black hole solutions.
Recently, non-circular stationary black hole solutions have been obtained by using the disformal transformation in DHOST theories~\cite{Anson:2020trg,BenAchour:2020fgy}, while the associated scalar fields have linear time dependence.

The aims of this paper are to explore the possibility of the existence of non-circular stationary black hole solutions 
in string-inspired gravity, and to develop the analysis methods for the construction of such solutions. 
To this end, we focus on a particular class of the quadratic gravity theories: dynamical Chern--Simons (dCS) gravity~\cite{Smith:2007jm,Alexander:2009tp} and shift symmetric Einstein scalar Gauss--Bonnet (ESGB) gravity~\cite{Sotiriou:2013qea,Sotiriou:2014pfa}. 
These two gravity theories modify general relativity by introducing dynamical scalar fields coupled with curvature invariants. 
In these quadratic gravity theories, 
stationary axisymmetric circular black hole solutions that differ from the Kerr spacetime have been found in slow rotation approximation~\cite{Yunes:2009hc,Yagi:2012ya,Pani:2011gy,Ayzenberg:2014aka} as well as non-perturbative numerical approaches~\cite{Delsate:2018ome,Delgado:2020rev}. 
We should note that these previous works assume stationarity of the scalar fields. 
In this paper, we consider perturbative solutions around the Schwarzschild spacetime and find slowly rotating black hole solutions up to quadratic order in the spin and linear order in coupling constants without assuming the circularity condition and stationarity of the scalar fields. 
We show that the stationarity of the effective stress energy tensors determines the time dependence of the scalar fields. From this condition, 
the linear time dependence of the scalar fields is derived. 
This result is natural because spacetimes can be stationary even if scalar fields have the linear time dependence due to the shift symmetry of the theories. 
We note that our analysis explicitly show the other time dependence of the scalar fields is impossible.
However, 
these time dependent scalar fields yield singular behavior of the metric functions at the black hole horizons in dCS gravity,\footnote{Note that the corresponding metric functions are non-circular but they are singular at the horizon. } 
and they are incompatible with the stationarity of the spacetime in ESGB gravity. 
Thus, these quadratic gravity theories admit only known solutions in~\cite{Yunes:2009hc,Yagi:2012ya,Pani:2011gy,Ayzenberg:2014aka} as regular stationary solutions. Our results suggest the non-existence of rotating non-circular black holes in these quadratic gravity theories.

This paper is organized as follows. 
In Sec.~\ref{sec:ThyandApp}, 
at the beginning, we review dCS gravity and ESGB gravity briefly. 
After that, we explain the approximation scheme. 
In Secs.~\ref{sec:dCSgravity} and~\ref{sec:ESGBgravity}, 
we study slowly rotating black holes in dCS gravity and ESGB gravity, respectively. 
Section~\ref{sec:conclusion} is devoted to summary and discussion. 
In Appendix~\ref{app:circularity}, 
we review the circularity condition. 
In Appendix~\ref{appB}, 
we review the linear stationary metric perturbation in the Regge--Wheeler gauge. 
In Appendix~\ref{appC}, 
we discuss the homogeneous solutions for the Einstein equations. 
In what follows, we use the geometric units: $c=G=1$ and parentheses in index lists for symmetrization:  $A_{(\mu \nu)}=(A_{\mu \nu}+A_{\nu\mu})/2$.

\section{Theories and approximation schemes}
\label{sec:ThyandApp}

\subsection{dynamical Chern--Simons gravity and shift symmetric Einstein scalar Gauss--Bonnet gravity}

In this paper, we focus on quadratic gravity theories, which introduce dynamical scalar fields coupling to curvature squared terms. 
In particular, 
we consider dynamical Chern--Simons (dCS) gravity and shift symmetric Einstein scalar Gauss--Bonnet (ESGB) gravity. 
The actions are given by
\begin{align}
    S_{\mathrm{dCS}} &= \int \mathrm{d} x^4 \sqrt{-g} \left [
    \kappa R + \frac{\alpha}{4} \vartheta R_{\nu \mu \rho \sigma} \widetilde R^{\mu \nu \rho \sigma} - \frac{\beta}{2} \{
    \nabla _{\mu} \vartheta \nabla ^{\mu} \vartheta + 2 V_{\mathrm{dCS}}(\vartheta)
    \}
     \right ],
     \\
     S_{\mathrm{ESGB}} &= \int \mathrm{d} x^4 \sqrt{-g} 
     \left [
    \kappa R + \alpha  \varphi \{ R^2 - 4R_{\mu \nu}R^{\mu \nu} + R_{\mu \nu \rho \lambda} R^{\mu \nu \rho \lambda}\} - \frac{\beta}{2} \{
    \nabla _{\mu} \varphi \nabla ^{\mu} \varphi + 2 V_{\mathrm{ESGB}}(\varphi)
    \}
     \right ],
\end{align}
respectively. 
Here, $\kappa = 1/(16 \pi)$, $g$ denotes the determinant of $g_{\mu \nu}$, and $\widetilde R_{\mu \nu \rho \sigma}$ is the dual of the Riemann tensor defined by
\begin{align}
    \widetilde R^{\mu \nu \rho \sigma} = \frac{1}{2} \epsilon ^{\rho \sigma \alpha \beta } R^{\mu \nu} \, _{ \alpha \beta},
\end{align}
where $\epsilon ^{\mu \nu \rho \sigma}$ is the Levi--Civita tensor.
$\vartheta$ and $\varphi$ are scalar fields, and $V_{\mathrm{dCS}}(\vartheta)$ and $V_{\mathrm{ESGB}}(\varphi)$ are potentials of the each gravity theory, while $\alpha$ and $\beta$ are coupling constants. 
In this paper, we set $V_{\mathrm{dCS}}(\vartheta) = V_{\mathrm{ESGB}}(\varphi) = 0$ which implies that those theories possess shift symmetry, i.e., theories are invariant under the transformation $\vartheta \to \vartheta + \mathrm{const}.$ or $\varphi \to \varphi + \mathrm{const}.$
We take $\vartheta$, $\varphi$, and $\beta$ are dimensionless and $\alpha$ have dimensions of $(\mathrm{length})^2$. 
Hereafter, we set $\beta=1$.

\subsection{approximation schemes}

We consider the slow rotation and the weak coupling approximation.
The expansion parameters are the slow rotation parameter $\varepsilon (\ll 1) $ and the weak coupling parameter $\zeta (\ll 1)$ which are respectively defined by
\begin{align}
    \varepsilon \equiv \frac{a}{M}, \,\,\,
    \zeta \equiv \frac{\alpha^2}{\kappa M^4},
\end{align}
where $M$ is the mass of the system and $a$ is the spin parameter of the system defined as the angular momentum divided by $M$. 
In this paper, 
we study the slowly rotating solutions up to quadratic order in the spin and linear order in coupling constant.
The circular black hole solutions in dCS gravity theory have been discussed at linear order in the spin \cite{Yunes:2009hc} and quadratic order in the spin \cite{Yagi:2012ya}.
The non-perturbative spinning solution have been also obtained numerically in \cite{Delsate:2018ome}.

We employ the approximation schemes used in~\cite{Yunes:2009hc,Yagi:2012ya}.
We expand the metric in terms of the weak coupling parameter as
\begin{align}
    g_{\mu \nu} = g_{\mu \nu} ^{(0)} + \zeta g_{\mu \nu} ^{(1)} + \mathcal{O} (\zeta ^2),
    \label{eq:expandzeta}
\end{align}
where $g_{\mu \nu} ^{(0)}$ is the exact Kerr metric
\footnote{We note that introducing small correction terms to the Einstein-Hilbert action, 
the theory may admit solutions which are very different from the case of general relativity (see an example in higher dimensional Einstein Gauss-Bonnet gravity \cite{Boulware:1985wk, Deser:2005jf}). 
In this paper, we focus on solutions which become those of general relativity in $\zeta \to 0$ limit.}
, which is expressed in the Boyer-Lindquist coordinates $\{ t, r, \theta, \phi \}$:
\begin{align}
    \mathrm{d}s^2 = - \frac{\Sigma \Delta}{A} \mathrm{d}t^2
    + \frac{\Sigma}{\Delta} \mathrm{d}r^2
    + \Sigma \mathrm{d} \theta ^2
    + \frac{A}{\Sigma} \sin ^2 \theta \left [ \mathrm{d} \phi - \frac{a(r^2 + a^2 - \Delta)}{A} \mathrm{d}t \right]^2,
    \label{eq:kerrmetric}
\end{align}
with
\begin{align}
    \Sigma &\equiv r^2 + a^2 \cos^2 \theta,
    \\
    \Delta &\equiv r^2 + a^2 - 2Mr,
    \\
    A &\equiv (r^2 + a^2)^2 - \Delta a^2 \sin ^2 \theta.
\end{align}
Next, we re-expand the metric~\eqref{eq:expandzeta} in terms of the slow rotation parameter as 
\begin{align}
   g_{\mu \nu} ^{(0)} &= g_{\mu \nu} ^{(0,0)} + \varepsilon g_{\mu \nu} ^{(1,0)} + \varepsilon ^2 g_{\mu \nu}^{(2,0)} + \mathcal{O} (\varepsilon ^3),
   \\
   \zeta g_{\mu \nu}^{(1)} &= \zeta g_{\mu \nu} ^{(0,1)} + \varepsilon \zeta g_{\mu \nu}^{(1,1)} + \varepsilon ^2 \zeta g_{\mu \nu} ^{(2,1)} + \mathcal{O}(\varepsilon ^3 \zeta).
\end{align}
The metric function $g_{\mu \nu} ^{(0,0)}, g_{\mu \nu} ^{(1,0)},$ and $g_{\mu \nu}^{(2,0)}$ are obtained from the expansion of the Kerr metric~\eqref{eq:kerrmetric} in terms of the slow rotation parameter up to quadratic order.
Note that $g_{\mu \nu} ^ {(m,n)} \propto \varepsilon ^m \zeta ^n$. We can express the metric function $g_{\mu \nu}^{(m,n)}$ in the Regge--Wheeler gauge\footnote{While we are interested in non-linear perturbations, at each order, the highest order corrections can be treated as linear perturbations with source terms from the lower order corrections. 
Thus, we can use the form of metric functions in Appendix~\ref{appB} at each order.
We note that the definition of the functions $H_{0\ell}^{(m,n)}$ and $H_{2\ell}^{(m,n)}$ in Eq.~\eqref{eq:perturbation} are slightly different from those in Appendix~\ref{appB}.
}
\begin{align}
     g_{\mu \nu} ^{(m,n)} = 
     \sum_{\ell = 0} ^{\infty}
     \left(
    \begin{array}{cccc}
      f(r) H_{0\ell}^{(m,n)}(r) & H_{1\ell}^{(m,n)}(r) & 0 & h_{0\ell}^{(m,n)}(r)\sin \theta \partial _{\theta} \\
        & H_{2\ell}^{(m,n)}(r)/f(r) & 0 & h_{1\ell}^{(m,n)}(r)\sin \theta \partial _{\theta} \\
       \mathrm{symm. } &   & r^2 K_{\ell}^{(m,n)}(r) & 0 \\
       & & & r^2 \sin ^2\theta K_{\ell}^{(m,n)}(r)
    \end{array}
  \right) Y_{\ell 0 }(\theta),
  \label{eq:perturbation}
\end{align}
where $Y_{\ell 0}$ is the spherical harmonics with vanishing azimuthal number. 
Note that we consider only $Y_{\ell 0}$ perturbation because we assume the axisymmetry of the system. 
In a similar way, we expand $\vartheta$ and $\varphi$ as follows
\begin{align}
    \vartheta &= \alpha \left[
    \vartheta^{(0,\frac{1}{2})} +
    \varepsilon   \vartheta^{(1,\frac{1}{2})} +
    \varepsilon^2   \vartheta^{(2,\frac{1}{2})}
    \right]
    + \mathcal{O} (\varepsilon ^3 \alpha ),
    \\
    \varphi &= \alpha \left[
    \varphi^{(0,\frac{1}{2})} +
    \varepsilon \varphi^{(1,\frac{1}{2})} +
    \varepsilon^{2} \varphi^{(2,\frac{1}{2})}
    \right] + \mathcal{O}(\varepsilon^3 \alpha).
\end{align}
Because the source terms of the equation of motion for scalar fields are the order $\mathcal{O}(\alpha)$ (see Eqs.~\eqref{eq:eomtheta} and~\eqref{eq:eomphi}), the leading terms of the scalar fields also are proportional to $\alpha$ in the above equations. Also, because ${\cal O}(\alpha) = {\cal O}(\zeta^{1/2})$,
we labeled the order of $\zeta$ on the scalar fields as $1/2$.
Here, we list up the scalar fields that can affect to the metric solution up to $\mathcal{O}(\varepsilon^2 \zeta)$.

\section{slowly rotating black holes in dCS gravity}
\label{sec:dCSgravity}

For dCS gravity, the field equations are given by
\begin{align}
    G_{\mu \nu} 
    & = 
    \frac{1}{2\kappa} T^{\vartheta}_{\mu \nu}
    - \frac{\alpha}{\kappa} \mathcal{C}_{\mu \nu} 
    \equiv T^{\mathrm{dCS}}_{\mu \nu},
    \label{eq:fieldeq}
    \\
    \mathcal{C}^{\mu \nu} 
    &\equiv 
    (\nabla _{\sigma} \vartheta) \epsilon ^{\sigma \delta \alpha ( \mu} R^{\nu )}\,_{\delta} + (\nabla _{\sigma} \nabla _{\delta} \vartheta) \widetilde R ^{\delta (\mu \nu) \sigma},
    \label{eq:ctensor}
    \\
    T^{\vartheta}_{\mu \nu} 
    &\equiv  
    (\nabla_{\mu} \vartheta)(\nabla_{\nu} \vartheta) - \frac{1}{2} g_{\mu \nu}  (\nabla_{\delta} \vartheta)
    (\nabla ^{\delta} \vartheta),
    \\
    \Box \vartheta 
    &= 
    - \frac{\alpha}{4} R_{\nu \mu \rho \sigma} \widetilde R^{\mu \nu \rho \sigma}. 
    \label{eq:eomtheta}
\end{align}
In Eq.~\eqref{eq:eomtheta}, 
we can expand the source terms by only spin parameter as follows 
\begin{align}
    \alpha R_{\nu \mu \rho \sigma} \widetilde R^{\mu \nu \rho \sigma}
    = 
    \alpha
    \left(
    \frac{288 M^3 \varepsilon \cos{\theta}}{r^7}
    \left( 1 - \frac{28M^2}{3 r^2} \varepsilon^2 \cos^2 \theta  \right)
    +\mathcal{O}(\varepsilon ^5)
    \right) 
    + \mathcal{O}(\alpha^3),
\end{align}
where we used that the metric is expanded by $\zeta$ as in Eq.~\eqref{eq:expandzeta}.
Notice that the source term contains only odd power of $\varepsilon$.

We assume the background spherically symmetric spacetime is the Schwarzschild spacetime. 
Under this assumption, 
the only second term of the $\mathcal{C}$-tensor~\eqref{eq:ctensor} is relevant for the metric solution to $\mathcal{O}(\zeta)$: 
\begin{align}
  \frac{\alpha}{\kappa}\mathcal{C}_{\mu \nu}  
  =
  \frac{\alpha}{\kappa}(\nabla ^{\sigma} \nabla ^{\delta}\, \vartheta) \widetilde R _{\delta (\mu \nu) \sigma} + \mathcal{O}(\zeta ^2). 
\end{align}
We expand the scalar filed with the spherical harmonics as $\vartheta^{(m,\frac{1}{2})} = \sum_{\ell}\Theta^{(m,\frac{1}{2})}_{\ell}(t,r) Y_{\ell 0}(\theta)$. 
In the following subsections, 
we see that the time dependence of $\vartheta$ is determined from the stationarity of the effective stress energy tensor $T_{\mu \nu}^{\mathrm{dCS}}$:
\begin{align}
    \partial_t T_{\mu \nu}^{\mathrm{dCS}}=0.
    \label{eq:stationarycondcs}
\end{align}
In addition, 
we impose the following condition 
\begin{align}
    T_{ti}^{\mathrm{dCS}} |_{\ell=0}=0,
    \label{eq:Tdcsti}
\end{align}
where $i=r,\theta$, and $\phi$, because these components of the Einstein tensor vanish for stationary spacetimes for $\ell=0$.

\subsection{$\mathcal{O}(\zeta)$ corrections}

Supposing that in the limit $\varepsilon \to 0$, 
the spacetime possesses the spherical symmetry, 
we need to consider only $\ell=0$ mode.\footnote{In dCS gravity, the static uniqueness of the metric with stationary scalar fields is proven in~\cite{Shiromizu:2013pna}. 
}
Thus, we can express $\vartheta ^{(0,\frac{1}{2})} (t,r) = \Theta_{0}^{(0,\frac{1}{2})}(t,r) Y_{00}$. 
Since on any spherical symmetric spacetime, 
the $\mathcal{C}$-tensor with $\ell=0$ mode is identically zero, 
the effective stress energy tensor contains only $T_{\mu \nu}^{\vartheta}$. 
The $tr$-component of the stationary condition~\eqref{eq:stationarycondcs} becomes  
\begin{align}
    \partial_t\Theta_{0}^{(0,\frac{1}{2})}
    \,\partial_t \partial_r \Theta_{0}^{(0,\frac{1}{2})} + \partial_r \Theta_{0}^{(0,\frac{1}{2})} \partial^2 _t \Theta_{0}^{(0,\frac{1}{2})} = 0.
\end{align}
The general solution is given by
\begin{align}
    \Theta_{0}^{(0,\frac{1}{2})}(t,r) 
    =
    \mathcal{F}_{0}^{(0,\frac{1}{2})}(r) + t\, \mathcal{G}_{0}^{(0,\frac{1}{2})}(r) + \mathcal{H}_{0}^{(0,\frac{1}{2})}(t),
\end{align}
where $\mathcal{F}_{0}^{(0,\frac{1}{2})}$ and $\mathcal{G}_{0}^{(0,\frac{1}{2})}$ are arbitrary functions of $r$, while $\mathcal{H}_{0}^{(0,\frac{1}{2})}$ is an arbitrary function of $t$. 
After substituting this into Eq.~\eqref{eq:stationarycondcs}, 
we solve $tt$-component and $tr$-component of Eq.~\eqref{eq:stationarycondcs} simultaneously with respect to $\mathrm{d} \mathcal{H}_{0}^{(0,\frac{1}{2})} / \mathrm{d} t$ and $\mathrm{d}^2 \mathcal{H}_{0}^{(0,\frac{1}{2})} / \mathrm{d} t^2$, then we have 
\begin{align}
    \frac{\mathrm{d}^2 \mathcal{H}_{0}^{(0,\frac{1}{2})}(t)}{\mathrm{d} t^2}  
    = \left ( 1 - \frac{2M}{r} \right) \frac{\mathrm{d} \mathcal{G}_{0}^{(0,\frac{1}{2})}(r)}{\mathrm{d} r}.
\end{align}
Since the left hand side is a function of only $t$ and the right hand side is a function of only $r$, 
both sides have to be equal to a constant $\mathtt{d}_{0}^{(0,\frac{1}{2})}$. 
We obtain the expressions for $\mathcal{G}_{0}^{(0,\frac{1}{2})}$ and $\mathcal{H}_{0}^{(0,\frac{1}{2})}$ as follows
\begin{align}
    \mathcal{G}_{0}^{(0,\frac{1}{2})}(r) 
    &= 
    \mathtt{a}_{0}^{(0,\frac{1}{2})} 
    + \mathtt{d}_{0}^{(0,\frac{1}{2})} [r + \log (r-2M)],
    \\
    \mathcal{H}_{0}^{(0,\frac{1}{2})}(t) 
    &= 
    \frac{1}{2} \mathtt{d}_{0}^{(0,\frac{1}{2})} t^2 + \mathtt{b}_{0}^{(0,\frac{1}{2})} t + \mathtt{c}_{0}^{(0,\frac{1}{2})}, 
\end{align}
where $\mathtt{a}_{0}^{(0,\frac{1}{2})}$, $\mathtt{b}_{0}^{(0,\frac{1}{2})}$, and $\mathtt{c}_{0}^{(0,\frac{1}{2})}$ are integration constants. 
Substituting $\mathcal{G}_{0}^{(0,\frac{1}{2})}$ and $\mathcal{H}_{0}^{(0,\frac{1}{2})}$ into Eq.~\eqref{eq:stationarycondcs}, 
we obtain $\mathtt{d}_{0}^{(0,\frac{1}{2})}=0$. 
Consequently, we obtain the form of $\Theta_{0}^{(0,\frac{1}{2})}$
\footnote{
Solving the equation of motion~\eqref{eq:eomtheta} with~\eqref{eq:Theta01_20}, we obtain the configuration of the scalar field which is regular at $r=2M$: $\vartheta ^{(0,\frac{1}{2})} \sim q_{0} v$ where $v$ is an advanced null coordinate expressed as $v \sim t + 2M \log(r-2M)$. 
However, we need to set $q_0=0$ to satisfy the condition~\eqref{eq:Tdcsti}. }
:
\begin{align}
    \Theta_{0}^{(0,\frac{1}{2})}(t,r) = 
    \mathcal{F}_{0}^{(0,\frac{1}{2})}(r) +
    (\mathtt{a}_{0}^{(0,\frac{1}{2})} + \mathtt{b}_{0}^{(0,\frac{1}{2})}) t + \mathtt{c}_{0}^{(0,\frac{1}{2})}. 
    \label{eq:Theta01_20}
\end{align}
Substituting this into Eq.~\eqref{eq:Tdcsti}, 
from the $tr$-component, we have 
\begin{align}
    \frac{\mathrm{d}\mathcal{F}_{0}^{(0,\frac{1}{2})}(r)}{\mathrm{d}r}
    \left( \mathtt{a}_{0}^{(0,\frac{1}{2})} + \mathtt{b}_{0}^{(0,\frac{1}{2})} \right)
    =0. 
\end{align}
There are two branches to satisfy this equation: $\mathcal{F}_{0}^{(0,\frac{1}{2})}(r) = \mathrm{const.}$ or $ \mathtt{a}_{0}^{(0,\frac{1}{2})} =- \mathtt{b}_{0}^{(0,\frac{1}{2})}$. 
We choose the former branch and then the scalar field depends $t$ only 
\begin{align}
    \vartheta^{(0,\frac{1}{2})} (t) 
    = 
    q_0 t + \mathrm{const.},
    \label{eq:vartheta1_20}
\end{align}
where $q_0(= [\mathtt{a}_{0}^{(0,\frac{1}{2})} + \mathtt{b}_{0}^{(0,\frac{1}{2})}]Y_{00})$. 
The expression for the scalar field~\eqref{eq:vartheta1_20} is a solution of the equation of motion~\eqref{eq:eomtheta}. 
We note that the scalar field $\vartheta ^{(0,\frac{1}{2})}$ becomes stationary when we choose the later branch. 
Solving the equation of motion~\eqref{eq:eomtheta} with this stationary scalar field configuration and imposing the regularity condition at $r=2M$, 
the scalar field becomes trivial: $\vartheta^{(0,\frac{1}{2})} = \mathrm{const.}$

Let us move on to the discuss on the metric. 
For $\ell=0$ mode, the non-vanishing components of the metric perturbation are $H_{00}^{(0,1)}$ and $H_{20}^{(0,1)}$. 
We solve the $tt$- and $rr$-components of the field equations~\eqref{eq:fieldeq} to find $H_{00}^{(0,1)}$ and $H_{20}^{(0,1)}$: 
\begin{align}
    H_{00}^{(0,1)}(r) 
    &= 
    -\frac{M^4 \sqrt{\pi}}{3} q_{0}^{2} r^2 \left[ 1 + \frac{15M}{2r} - \frac{84M^3}{r^3} \log(r-2M) - \frac{6M^2}{r^2} (7 - 6 \log(r-2M)) \right] - \frac{c_1^{(0,1)}}{r f} + c_2^{(0,1)},
    \\
    H_{20}^{(0,1)}(r)
    &=
    \frac{M^4 \sqrt{\pi}}{6} q_{0}^2 r^2 
    \left[
    1 + \frac{3M}{r} + \frac{12M^2}{r^2} + \frac{24M^3}{r^3} \log(r-2M)
    \right] - \frac{c_1^{(0,1)}}{r f},
\end{align}
where $c_1^{(0,1)}$ and $c_2^{(0,1)}$ are integration constants. 
The logarithmic dependence in $H_{00}^{(0,1)}$ and $H_{20}^{(0,1)}$ leads to the divergence of the  Kretschmann invariant $R_{\mu \nu \rho \lambda}R^{\mu \nu \rho \lambda}$ at $r=2M$. 
In order to remove this divergence, 
we need to set $q_{0}=0$, 
which implies $\vartheta^{(0,\frac{1}{2})} = \mathrm{const}$. 
We set $c_1^{(0,1)}=c_2^{(0,1)}=0$ so that the metric to be asymptotically flat at the spatial infinity, and the mass of the system does not change from $M$.

\subsection{$\mathcal{O}(\varepsilon \zeta)$ corrections}

Because the stress energy tensor of the scalar field $T_{\mu \nu}^{\vartheta}$ is always of order of $\mathcal{O}(\varepsilon ^2 \zeta)$, $T_{\mu \nu}^{\mathrm{dCS}}$ contains only $\mathcal{C}$-tensor~\eqref{eq:ctensor}. 
At the order $\mathcal{O}(\varepsilon \zeta)$, 
$\mathcal{C}$-tensor is identically zero for $\ell=0$ mode. 
This means that we cannot obtain the form of the scalar field with $\ell=0$ from the stationary condition~\eqref{eq:stationarycondcs}.
As shown later, the time dependence of $\vartheta^{(1,\frac{1}{2})}$ for $\ell=0$ is determined by the stationary condition of the order $\mathcal{O}(\varepsilon ^2 \zeta)$.

For $\ell \ge 1$, the $\theta \theta$-component of the stationary condition~\eqref{eq:stationarycondcs} becomes $\partial_t ^2 \Theta^{(1,\frac{1}{2})}_{\ell}(t,r)=0$. 
The general solution is given by
\begin{align}
   \Theta^{(1,\frac{1}{2})}_{\ell}(t,r) = \mathcal{F}^{(1,\frac{1}{2})}_{\ell}(r) + t\, \mathcal{G}^{(1,\frac{1}{2})}_{\ell}(r).
\end{align}
Substituting this into Eq.~\eqref{eq:stationarycondcs}, we obtain
\begin{align}
  r\, \frac{\mathrm{d}\mathcal{G}^{(1,\frac{1}{2})}_{\ell}(r)}{\mathrm{d}r} -  \mathcal{G}^{(1,\frac{1}{2})}_{\ell}(r) = 0.
\end{align}
The general solution is 
\begin{align}
 \mathcal{G}^{(1,\frac{1}{2})}_{\ell}(r) = \mathtt{a}^{(1,\frac{1}{2})}_{\ell} \,r,
\end{align}
where $\mathtt{a^{(1,\frac{1}{2})}_{\ell}}$ is an integration constant. 
We consider the cases of $\ell=1$ and $\ell \ge 2$, separately. 
For $\ell = 1$, the equation of motion for the scalar field~\eqref{eq:eomtheta} can be written as 
\begin{align}
    \mathtt{a}^{(1,\frac{1}{2})}_{1}\, t = \frac{24 M^3 \sqrt{3 \pi}}{r^5} - \mathcal{F}^{(1,\frac{1}{2})}_{1}(r) + (r-M)\frac{\mathrm{d}\mathcal{F}^{(1,\frac{1}{2})}_{1}(r)}{\mathrm{d}r} + \frac{2(r-2M)}{2} \frac{\mathrm{d}^2 \mathcal{F}^{(1,\frac{1}{2})}_{1}(r)}{\mathrm{d}r^2},
    \label{eq:c1eqoml1}
\end{align}
which means $\mathtt{a}^{(1,\frac{1}{2})}_{1}=0$ so that the above equation holds for all $t$ and $r$.
By solving Eq.~\eqref{eq:c1eqoml1}, 
we obtain the form of $\mathcal{F}^{(1,\frac{1}{2})}_{1}$:
\begin{align}
    \mathcal{F}^{(1,\frac{1}{2})}_{1}(r) = 
    \frac{5\sqrt{3 \pi}}{12r^2} 
    \left( 
    1 + \frac{2M}{r} + \frac{18M^2}{5r^2} 
    \right).
\end{align}
We chose the integration constants as the scalar field is regular everywhere. 
For $\ell \ge 2$, 
we have $\mathtt{a}^{(1,\frac{1}{2})}_{\ell}=0$ from the similar analysis of the equation of motion for the scalar field~\eqref{eq:eomtheta}. 
Then, Eq.~\eqref{eq:eomtheta} reduces to  
\begin{align}
   \ell (\ell + 1) \mathcal{F}^{(1,\frac{1}{2})}_{\ell}(x) - (2x+1) \frac{\mathrm{d} \mathcal{F}^{(1,\frac{1}{2})}_{\ell}(x)}{\mathrm{d} x} - x(x+1) \frac{\mathrm{d}^2 \mathcal{F}^{(1,\frac{1}{2})}_{\ell}(x)}{\mathrm{d}x^2} = 0, 
   \label{eq:Fequation}
\end{align}
where we introduce a new variable $x \equiv r/2M -1$. 
The general solution is given by 
\begin{align}
    \mathcal{F}^{(1,\frac{1}{2})}_{\ell}(x) = \mathtt{b}^{(1,\frac{1}{2})}_{\ell} P_{\ell}(1+2x) + \mathtt{c}^{(1,\frac{1}{2})}_{\ell}\, Q_{\ell}(1+2x),
    \label{eq;Fsolution}
\end{align}
where $\mathtt{b}^{(1,\frac{1}{2})}_{\ell}$ and $\mathtt{c}^{(1,\frac{1}{2})}_{\ell}$ are integration constants, and $P_{\ell}$ and $Q_{\ell}$ are Legendre polynomials of first kind and second kind, respectively. 
Requiring that the scalar field is regular everywhere, 
we should set $\mathtt{b}^{(1,\frac{1}{2})}_{\ell}=\mathtt{c}^{(1,\frac{1}{2})}_{\ell}=0$.\footnote{For large $x$, Legendre polynomial of first kind diverges as $P_\ell(1+2x) \sim x^\ell$. 
For small $x$, Legendre polynomial of second kind diverge because it  behaves as $Q_{\ell}(1+2x) \sim \log\left( (x+1)/x \right)$.} 
Therefore, the configuration of the scalar field $\vartheta^{(1,\frac{1}{2})}$ for $\ell \ge 1$ is given by 
\begin{align}
    \vartheta^{(1,\frac{1}{2})} (r,\theta) = \frac{5}{8} \alpha \varepsilon \frac{\cos \theta}{r^2} \left( 1 + \frac{2M}{r} + \frac{18M^2}{5r^2} \right).
\end{align}
We note again that the time dependence of $\vartheta^{(1,\frac{1}{2})}$ for $\ell=0$ is determined by the stationary condition of the order $\mathcal{O}(\varepsilon ^2 \zeta)$.

We derive the metric solution of the order $\mathcal{O}(\varepsilon \zeta)$. 
The the effective stress energy tensor has only non-trivial component $T_{t \phi}^{\mathrm{dCS}}$ for $\ell=1$. 
The $t\phi$-component of the field equation~\eqref{eq:fieldeq} for $\ell=1$ becomes
\begin{align}
    r^2 \frac{\mathrm{d}^2 h_{01}^{(1,1)}}{\mathrm{d}r^2} 
    - 2 \frac{\mathrm{d}h_{01}^{(1,1)}}{\mathrm{d}r} = - \frac{15\sqrt{3\pi} M^5}{2r^4}
    \left(
    1 + \frac{8M}{3r} + \frac{6M^2}{r^2}
    \right),
\end{align}
and the solution is given by 
\begin{align}
    h_{01}^{(1,1)}(r) = -\frac{5 \sqrt{3\pi} M^5}{4r^4}
    \left(
    1 + \frac{12M}{7r} + \frac{27M^2}{10r^2}
    \right)
    + 
    \frac{c_1^{(1,1)}}{r} 
    + 
    c_2^{(1,1)}r^2,
    \label{eq:oddh0111dcs}
\end{align}
where $c_1^{(1,1)}$ and $c_2^{(1,1)}$ are integration constants. 
We set $c_1^{(1,1)}=c_2^{(1,1)}=0$ so that the metric is asymptotically flat at the spatial infinity and the amplitude of the angular momentum does not change from $Ma$. 
For $\ell=0$ and $\ell \ge 2$, the field equations~\eqref{eq:fieldeq} are homogeneous equations, the solutions are discussed in Appendix~\ref{appC}. 
The metric perturbations for $\ell=0$ and  $\ell \ge 2$ vanish by imposing that the spacetime is asymptotically flat and regular at the horizon. 
The result~\eqref{eq:oddh0111dcs} with $c_1^{(1,1)}$ = $c_2^{(1,1)} = 0$ coincides with the metric solution obtained in~\cite{Yunes:2009hc}. 
Note that the spacetime is circular up to this order.

\subsection{$\mathcal{O}(\varepsilon^2 \zeta)$ corrections}

For $\ell=0$ mode, 
the $t\theta$-component of the stationary condition~\eqref{eq:stationarycondcs} is
\begin{align}
    2M \partial^2_t\Theta_{0}^{(1,\frac{1}{2})}(t,r) + 
    r(r-2M) \partial^2_t \partial_r \Theta_{0}^{(1,\frac{1}{2})}(t,r) = 0.
\end{align}
The general solution is given by 
\begin{align}
    \Theta_{0}^{(1,\frac{1}{2})}(t,r) 
    =  
    \mathcal{F}_{0}^{(1,\frac{1}{2})}(r) 
    +
    t\, \mathcal{G}_{0}^{(1,\frac{1}{2})}(r)
    + 
    \frac{r}{r-2M} \mathcal{H}_{0}^{(1,\frac{1}{2})}(t).
\end{align}
Substituting this into Eq.~\eqref{eq:stationarycondcs}, 
we solve the $r \theta$-component with respect to $\mathrm{d} \mathcal{G}_{0}^{(1,\frac{1}{2})} /\mathrm{d}r$: 
\begin{align}
    \frac{\mathrm{d}\mathcal{G}_{0}^{(1,\frac{1}{2})}(r)}{\mathrm{d}r} 
    = 
    \frac{1}{(r-2M)^2} 
    \left ( 
    2M \frac{\mathrm{d}\mathcal{H}_{0}^{(1,\frac{1}{2})}(t)}{\mathrm{d} t} - r^3 \frac{\mathrm{d}^3 \mathcal{H}_{0}^{(1,\frac{1}{2})}(t)}{\mathrm{d}t^3} 
    \right). 
    \label{eq:derivativec3}
\end{align}
Since the left hand side is a function of $r$, 
the right hand side also has to be a function of $r$. From that condition, we have 
\begin{align}
    \mathcal{H}_{0}^{(1,\frac{1}{2})}(t) 
    = 
    \mathtt{a}_{0}^{(1,\frac{1}{2})} 
    +
    \mathtt{b}_{0}^{(1,\frac{1}{2})} t,
\end{align}
where $\mathtt{a}_{0}^{(1,\frac{1}{2})}$ and $\mathtt{b}_{0}^{(1,\frac{1}{2})}$ are constants. 
Then we can integrate Eq.~\eqref{eq:derivativec3} to obtain the expression of $\mathcal{G}_{0}^{(1,\frac{1}{2})}$: 
\begin{align}
    \mathcal{G}_{0}^{(1,\frac{1}{2})}(r) = - \frac{2M \mathtt{b}_{0}^{(1,\frac{1}{2})}}{r-2M} + \mathtt{c}_{0}^{(1,\frac{1}{2})},
\end{align}
where $\mathtt{c}_{0}^{(1,\frac{1}{2})}$ is an integration constant. 
Thus, we have 
\begin{align}
    \Theta_{0}^{(1,\frac{1}{2})}(t,r) 
    = 
    \mathcal{F}_{0}^{(1,\frac{1}{2})}(r)
    +
    \frac{\mathtt{a}_{0}^{(1,\frac{1}{2})} r}{r-2M}
    +
    \left( \mathtt{b}_{0}^{(1,\frac{1}{2})} + \mathtt{c}_{0}^{(1,\frac{1}{2})} \right) t.
\end{align}
{}From the regularity of the scalar field at $r=2M$, we need to set $\mathtt{a}_{0}^{(1,\frac{1}{2})}=0$. 
From the $tr$-component of Eq.~\eqref{eq:Tdcsti}, 
we have
\begin{align}
    \frac{\mathrm{d}\mathcal{F}_{0}^{(1,\frac{1}{2})}(r)}{\mathrm{d}r}
    \left( \mathtt{b}_{0}^{(1,\frac{1}{2})} + \mathtt{c}_{0}^{(1,\frac{1}{2})} \right)
    =0. 
\end{align}
We choose the branch in which $\mathcal{F}_{0}^{(1,\frac{1}{2})} = \mathrm{const. }$ for the same reason as for the case of $\vartheta ^{(0,\frac{1}{2})}$, i.e, 
the other branch in which $\mathtt{b}_{0}^{(1,\frac{1}{2})} = - \mathtt{c}_{0}^{(1,\frac{1}{2})}$ makes the $\ell=0$ mode of the scalar field $\vartheta ^{(1,\frac{1}{2})}$ trivial: $\vartheta ^{(1,\frac{1}{2})}|_{\ell=0}=\mathrm{const.}$
Therefore, the scalar field $\vartheta^{(1,\frac{1}{2})}$ is given by 
\begin{align}
    \vartheta^{(1,\frac{1}{2})} (t,r,\theta) 
    =
    \alpha \varepsilon 
    \left[
    q\, t 
    +
    \frac{5 \cos \theta}{8r^2} \left(  1 + \frac{2M}{r} + \frac{18M^2}{5r^2} \right)
    +
    \mathrm{const}.
    \right],
    \label{eq:thetaqt}
\end{align}
where $q(=[\mathtt{b}_{0}^{(1,\frac{1}{2})} + \mathtt{c}_{0}^{(1,\frac{1}{2})}]Y_{00})$ is a constant.

We discuss the metric function $g_{\mu \nu}^{(2,1)}$. 
First, we focus on the non-circular component $H_{1 \ell}^{(2,1)}$. From the $tr$-component of the field equation~\eqref{eq:fieldeq} for $\ell =1$, we obtain $H_{11}^{(2,1)}$ as follows
\begin{align}
    H_{11}^{(2,1)}(r) = \frac{5 M^4 \sqrt{3 \pi} q}{12 f r} 
    \left( 1 + \frac{M}{r} + \frac{6M^2}{5r^2} - \frac{24 M^3}{5 r^3} \right). 
\end{align}
Since $H_{11}^{(2,1)}$ is proportional to $q$, 
we can see that the non-circularity comes from the linear time dependence of the scalar field $\vartheta^{(1,\frac{1}{2})}$. 
For $\ell \ge 2$, the equations for $H_{1\ell}^{(2,1)}$ become homogeneous, and 
these can be solved as $H_{1\ell}^{(2,1)} = 0$. We note that $H_{1\ell}^{(2,1)} = 0$ for $\ell = 0$ from the gauge condition.

Next, we discuss the other metric components. 
For $\ell =0$, we solve $tt$-component of the field equation~\eqref{eq:fieldeq}, which is a first order ordinary differential equation for $H_{20}^{(2,1)}$ and use $H_{20}^{(2,1)}$ to obtain $H_{00}^{(2,1)}$. 
The components $H_{00}^{(2,1)}$ and $H_{20}^{(2,1)}$ are given by
\begin{align}
    H_{00}^{(2,1)}(r) &=
    -\frac{5 \sqrt{\pi}}{192} \varepsilon ^{2} \zeta \frac{M^{5}}{r^{5} f}
    \left(
    1 
    + \frac{100M}{r} 
    + \frac{194M^{2}}{r^{2}} 
    + \frac{2220 M^{3}}{7 r^{3}} 
    - \frac{1512 M^{4}}{5 r^{4}}\right) 
    - \frac{c_1^{(2,1)}}{rf}
    + c_2^{(2,1)}
    \notag\\
    &- \frac{M^4 q^2 \sqrt{\pi}}{3 f} r^2 \varepsilon ^2 \zeta \left[
    1
    + \frac{15 M}{2 r}
    - \left( 19 - 36 \log (r-2M) \right) \frac{M^2}{r^2}
    - \left( 46 + 84 \log (r-2M) \right) \frac{M^3}{r^3}
    \right],
    \\
    H_{20}^{(2,1)}(r) &=
    -\frac{25 \sqrt{\pi}}{192}  \varepsilon^{2} \zeta \frac{M^{4}}{r^{4} f}
    \left(1
    + \frac{3 M}{r}
    + \frac{322 M^{2}}{5 r^{2}}
    + \frac{198 M^{3}}{5 r^{3}}
    + \frac{6276 M^{4}}{175 r^{4}}
    - \frac{17496 M^{5}}{25 r^{5}}
    \right) 
    - \frac{c_1^{(2,1)}}{rf}
    \notag\\
    & - \frac{M^4 q^2 \sqrt{\pi}}{6 f}r^2 \varepsilon ^2 \zeta
    \left[ 1
    + \frac{3 M}{r}
    + \frac{12 M^2}{r^2}
    + \frac{24 M^3}{r^3} \log (r-2M)
    \right], 
\end{align}
where $c_1^{(2,1)}$ and $c_2^{(2,1)}$ are integration constants. 
For $\ell =2$, there is no contribution from the time dependence of the scalar field because it appears only in $\ell=0$ mode (see Eq.~\eqref{eq:thetaqt}). 
Thus, $H_{02}^{(2,1)}$, $H_{22}^{(2,1)}$, and $K_{2}^{(2,1)}$ are given by the same forms obtained in~\cite{Yagi:2012ya}.\footnote{
The functions $H_{02}^{(2,1)}$, $H_{22}^{(2,1)}$, and $K_{2}^{(2,1)}$
correspond to the functions $H_{020}, H_{220}$ and $K_{20}$ in Eqs.(A18)-(A20) in~\cite{Yagi:2012ya}, respectively.}

Because the components $H_{00}^{(2,1)}$ and $H_{20}^{(2,1)}$ include the logarithmic divergence as $r \to 2M$, 
the non-circular spacetime obtained in this paper is singular at $r=2M$. 
Indeed, the Kretschmann invariant $R_{\mu \nu \rho \lambda} R^{\mu \nu \rho \lambda}$ diverges at $r=2M$. 
In dCS gravity, therefore, there does not exist the non-circular regular black hole solution at least around the Schwarzschild spacetime up to the order $\mathcal{O}(\varepsilon^2 \zeta)$.

\section{slowly rotating black holes in ESGB gravity}
\label{sec:ESGBgravity}

For ESGB gravity, the field equations are given by 
\begin{align}
    G_{\mu \nu} 
    &=  \frac{1}{2\kappa} T^{\varphi}_{\mu \nu}
    - \frac{\alpha}{\kappa} \mathcal{D}_{\mu \nu} 
    \equiv T^{\mathrm{ESGB}}_{\mu \nu},
    \label{fieldeqesgb}
    \\
    \mathcal{D}_{\mu \nu} 
    &\equiv -2 (\nabla_{\mu} \nabla_{\nu}\varphi) R 
    + 2(\nabla ^{\sigma} \nabla _{\sigma} \varphi) (g_{\mu \nu}R - 2R_{\mu \nu}) 
    \notag\\
    &\,\,\,\,\,\,\,
    + 8 R_{\sigma (\mu} \nabla^\sigma \nabla_{\nu )} \varphi
    - 4 g_{\mu \nu} (\nabla_{\sigma} \nabla_{\delta} \varphi) R^{\sigma \delta} 
    + 4 (\nabla^{\sigma} \nabla^{\delta} \varphi) R_{\mu \sigma \nu \delta} ,
    \label{eq:Dtensor}
    \\
    T^{\varphi}_{\mu \nu} 
    &\equiv  (\nabla_{\mu} \varphi)(\nabla_{\nu} \varphi) - \frac{1}{2} g_{\mu \nu}  (\nabla_{\delta} \varphi) (\nabla ^{\delta} \varphi), 
    \\
    \Box \varphi 
    &= - \alpha(R^2 -4 R_{\mu \nu}R^{\mu \nu} + R_{\mu \nu \rho \sigma} R^{\mu \nu \rho \sigma}).
    \label{eq:eomphi}
\end{align}
We consider the perturbed solution around the Schwarzschild spacetime. 
We can expand the source term of Eq.~\eqref{eq:eomphi} in terms of only the spin parameter as follows 
\begin{align}
     \alpha(R^2 -4 R_{\mu \nu}R^{\mu \nu} + R_{\mu \nu \rho \sigma} R^{\mu \nu \rho \sigma})
    = 
    \alpha
    \left(
    \frac{48 M^2}{r^6} 
    \left( 1 - \frac{21  M^2 }{r^2} \varepsilon ^2 \cos^2 \theta \right)
    +\mathcal{O}(\varepsilon ^4)
    \right)
    + \mathcal{O}(\alpha^3), 
\end{align}
where we used that the metric is expanded by $\zeta$ as in Eq.~\eqref{eq:expandzeta}. 
Note that the source term contains the even power of $\varepsilon$. 

When we focus on the perturbed solution around the Schwarzschild spacetime, 
only the last term of $\mathcal{D}$-tensor~\eqref{eq:Dtensor} is relevant for the linear order of $\zeta$:  
\begin{align}
    \frac{\alpha}{\kappa} \mathcal{D}_{\mu \nu} 
    =
    \frac{4 \alpha}{\kappa} (\nabla^{\sigma} \nabla^{\delta} \varphi) R_{\mu \sigma \nu \delta}  
    + 
    \mathcal{O}(\zeta^2). 
\end{align}
We expand the scalar field with the spherical harmonics as $\varphi^{(m,\frac{1}{2})} = \sum_{\ell} \Phi^{(m,\frac{1}{2})}_{\ell}(t,r) Y_{\ell0}(\theta) $. 
As we discussed in the previous section, 
we impose the stationary condition to the effective stress energy tensor $T_{\mu \nu}^{\mathrm{ESGB}}$: 
\begin{align}
    \partial _t T_{\mu \nu}^{\mathrm{ESGB}} = 0. 
    \label{eq:stationaryconesgb}
\end{align}
We also should impose 
\begin{align}
    T_{ti}^{\mathrm{ESGB}} |_{\ell=0}=0,
    \label{eq:Tti}
\end{align}
where $i=r,\theta$, and $\phi$, because these components of the Einstein tensor vanish for stationary spacetimes for $\ell=0$. 
While the stationary condition~\eqref{eq:stationaryconesgb} admits linear time dependence of the scalar fields, such behavior is prohibited from the condition~\eqref{eq:Tti} as shown in the following discussion.

\subsection{$\mathcal{O}(\zeta)$ corrections}

We assume that the spacetime possesses the spherical symmetry in the limit $\varepsilon \to 0$. This means that we can express $\varphi^{(0,\frac{1}{2})}(t,r) = \Phi^{(0,\frac{1}{2})}_{0}(t,r)Y_{00}$. 
Combining the $tr$-component of Eq.~\eqref{eq:Tti} and its time derivative, we obtain\footnote{
First we solve $tr$-component of Eq.~\eqref{eq:Tti} with respect to $\partial_r \Phi^{(0,1/2)}$,
and substitute it to the time derivative of the $tr$-component of Eq.~\eqref{eq:Tti}. Then, we can solve the differential equation and the general solution is Eq.~\eqref{eq:Phi01_20tr}.
}
\begin{align}
    \Phi^{(0,\frac{1}{2})}_{0}(t,r)
    = 
    \mathcal{I}^{(0,\frac{1}{2})}_{0}(r)
    + t \mathcal{J}^{(0,\frac{1}{2})}_{0}(r)
    + t^2 \mathcal{K}^{(0,\frac{1}{2})}_{0}(r)
    + \mathcal{L}^{(0,\frac{1}{2})}_{0}(t),
    \label{eq:Phi01_20tr}
\end{align}
where $\mathcal{I}^{(0,\frac{1}{2})}_{0}$, $\mathcal{J}^{(0,\frac{1}{2})}_{0}$, and $\mathcal{K}^{(0,\frac{1}{2})}_{0}$ are arbitrary functions of $r$, while $\mathcal{L}^{(0,\frac{1}{2})}_{0}$ is an arbitrary function of $t$. 
Substituting this expression into Eq.~\eqref{eq:Tti} again, 
we can solve the $tr$-component with respect to $\mathrm{d} \mathcal{L}^{(0,\frac{1}{2})}_{0}/\mathrm{d}t$ as 
\begin{align}
    \frac{\mathrm{d}\mathcal{L}^{(0,\frac{1}{2})}_{0}(t)}{\mathrm{d}t}
    = 
    -\frac{32M\sqrt{\pi}}{r^3 {\cal A}(t,r)}
    \left( 
    \frac{\mathrm{d}\mathcal{J}^{(0,\frac{1}{2})}_{0}(r)}{\mathrm{d}r} + 2t \frac{\mathrm{d}\mathcal{K}^{(0,\frac{1}{2})}_{0}(r)}{\mathrm{d}r}
    \right) 
    - 
    \left(
    \mathcal{J}^{(0,\frac{1}{2})}_{0}(r) + 2t \mathcal{K}^{(0,\frac{1}{2})}_{0}(r)
    \right),
    \label{eq:LL01_20t}
\end{align}
where
\begin{align}
    {\cal A}(t,r) \equiv \frac{\mathrm{d}\mathcal{I}^{(0,\frac{1}{2})}_{0}(r)}{\mathrm{d}r} + t \frac{\mathrm{d} \mathcal{J}^{(0,\frac{1}{2})}_{0}(r)}{\mathrm{d}r} + t^2 \frac{\mathrm{d}\mathcal{K}^{(0,\frac{1}{2})}_{0}(r)}{\mathrm{d}r}-\frac{32M^2\sqrt{\pi}}{r^4(r-2M)}.
\end{align}
{}From the condition for the right hand side of Eq.~\eqref{eq:LL01_20t} to be a function of only $t$, 
we obtain 
\begin{align}
    \mathcal{K}^{(0,\frac{1}{2})}_{0}(r) 
    &= \mathtt{f}^{(0,\frac{1}{2})}_{0},
    \\
    \mathcal{J}^{(0,\frac{1}{2})}_{0}(r) 
    &= \mathtt{g}^{(0,\frac{1}{2})}_{0},
\end{align}
where $\mathtt{f}^{(0,\frac{1}{2})}_{0}$ and $\mathtt{g}^{(0,\frac{1}{2})}_{0}$ are constants. 
Then, we integrate Eq.~\eqref{eq:LL01_20t} to obtain $\mathcal{L}^{(0,\frac{1}{2})}_{0}$: 
\begin{align}
    \mathcal{L}^{(0,\frac{1}{2})}_{0}(t) 
    = 
    - t^2\, \mathtt{f}^{(0,\frac{1}{2})}_{0} 
    - t\, \mathtt{g}^{(0,\frac{1}{2})}_{0} 
    + \mathtt{h}^{(0,\frac{1}{2})}_{0},
\end{align}
where $\mathtt{h}^{(0,\frac{1}{2})}_{0}$ is an integration constant. 
When we substitute this into Eq.~\eqref{eq:Phi01_20tr}, we obtain  
\begin{align}
    \Phi^{(0,\frac{1}{2})}_{0}(r) 
    = 
    \mathcal{I}^{(0,\frac{1}{2})}_{0}(r) + \mathrm{const}.
    \label{eq:phi01_2r}
\end{align}
Notice that the time dependence of the scalar field is not allowed. 

Substituting Eq.~\eqref{eq:phi01_2r} into the equation of motion for scalar field~\eqref{eq:eomphi} and impose the regularity condition at the horizon, 
we obtain the configuration of the scalar field $\varphi^{(0,\frac{1}{2})}$
\footnote{
We imposed the regularity condition at the horizon $r = 2 M$ to obtain this expression.
Here, we implicitly assumed the existence of the black hole, i.e., $M \neq 0$,
and we need to discuss $M = 0$ case separately.
In fact, $\varphi^{(0,\frac{1}{2})}(r) = 0$ for $M = 0$ case.
Thus, we cannot take a smooth $M \to 0$ limit to this solution.
}~\cite{Yunes:2011we}:
\begin{align}
    \varphi^{(0,\frac{1}{2})}(r) 
    = 
    \alpha \frac{2}{Mr} 
    \left( 
    1 + \frac{M}{r} + \frac{4M^2}{3r^2} 
    \right)
    + \mathrm{const}.
\end{align}

The metric of the order $\mathcal{O}(\zeta)$ can be obtained by solving the field equation~\eqref{fieldeqesgb}
\begin{align}
    g_{tt}^{(0,1)} (r)
    &= 
    - \frac{\zeta}{3} \frac{M^3}{r^3} 
    \left(
    1 + \frac{26 M}{r} + \frac{66M^2}{5r^2} + \frac{96M^3}{5r^3} - \frac{80M^4}{r^4}
    \right), 
    \\
    g_{rr}^{(0,1)} (r)
    &= 
    - \frac{\zeta}{f^2} \frac{M^2}{r^2}
    \left(
    1 + \frac{M}{r} + \frac{52M^2}{3r^2} + \frac{2M^3}{r^3} + \frac{16M^4}{5r^4} - \frac{368M^5}{3r^5}
    \right).
\end{align}
These expressions are same as those in~\cite{Yunes:2011we}.

\subsection{$\mathcal{O}(\varepsilon \zeta)$ corrections}

For $\ell=0$ mode, 
from the $tr$-component of Eq.~\eqref{eq:Tti}, 
we obtain 
\begin{align}
    \Phi^{(1,\frac{1}{2})}_{0}(t,r)
    = 
    \mathcal{I}^{(1,\frac{1}{2})}_{0}(r) 
    + 
    e^{\frac{r(r+4M)}{16M^2}}  \sqrt{r-2M}
    \mathcal{L}^{(1,\frac{1}{2})}_{0}(t),
\end{align}
where $\mathcal{I}^{(1,\frac{1}{2})}_{0}$ and $\mathcal{L}^{(1,\frac{1}{2})}_{0}$ are arbitrary functions of $r$ and $t$, respectively. 
From the $rr$-component and the $\theta \theta$-component of Eq.~\eqref{eq:Tti}, we obtain
\begin{align}
    \mathcal{L}^{(1,\frac{1}{2})}_{0} (t) 
    = 
    \mathtt{f}^{(1,\frac{1}{2})}_{0},
\end{align}
where $\mathtt{f}^{(1,\frac{1}{2})}_{0}$ is a constant. 
In order to remove the divergence of the scalar field at spatial infinity, we set $\mathtt{f}^{(1,\frac{1}{2})}_{0} = 0$, which means
\begin{align}
    \Phi^{(1,\frac{1}{2})}_{0}(r) = \mathcal{I}^{(1,\frac{1}{2})}_{0}(r)
    \label{eq:Phi11_20r}.
\end{align}
Substituting Eq.~\eqref{eq:Phi11_20r} into the equation of motion for the scalar field~\eqref{eq:eomphi}, and imposing the regularity at $r=2M$, we have
\begin{align}
    \Phi^{(1,\frac{1}{2})}_{0}(r)
    =
    \mathrm{const}.
\end{align}

For $\ell \ge 1$, 
the $t\theta$-component of the stationary condition~\eqref{eq:stationaryconesgb} is $\partial_t ^2 \Phi^{(1,\frac{1}{2})}_{\ell} = 0$. 
The general solution is given by 
\begin{align}
    \Phi^{(1,\frac{1}{2})}_{\ell}(t,r) 
    = 
    \mathcal{I}^{(1,\frac{1}{2})}_{\ell}(r)
    +
    t\, \mathcal{J}^{(1,\frac{1}{2})}_{\ell}(r),
\end{align}
where $\mathcal{I}^{(1,\frac{1}{2})}_{\ell}$ and $\mathcal{J}^{(1,\frac{1}{2})}_{\ell}$ are arbitrary functions of $r$. 
When we substitute this into Eq.~\eqref{eq:stationaryconesgb}, 
we obtain the expression for $\mathcal{J}^{(1,\frac{1}{2})}_{\ell}$:
\begin{align}
    \mathcal{J}^{(1,\frac{1}{2})}_{\ell}(r)
    = 
    \mathtt{f}^{(1,\frac{1}{2})}_{\ell}
    \left( 
    r - 3M
    \right)^{\ell(\ell + 1)/2},
\end{align}
where $\mathtt{f}^{(1,\frac{1}{2})}_{\ell}$ is an integration constant. 
Substituting this into Eq.~\eqref{eq:stationaryconesgb}, 
we obtain $\mathtt{f}^{(1,\frac{1}{2})}_{\ell}=0$. 
Since the equation of motion of the scalar field~\eqref{eq:eomphi} becomes the same form as~\eqref{eq:Fequation}, 
the expression for $\Phi^{(1,\frac{1}{2})}_{\ell}$ is given by
\begin{align}
    \Phi^{(1,\frac{1}{2})}_{\ell}(x) 
    = 
    \mathtt{g}^{(1,\frac{1}{2})}_{\ell} P_{\ell}(1+2x) 
    + 
    \mathtt{h}^{(1,\frac{1}{2})}_{\ell}
    Q_{\ell}(1+2x),
\end{align}
where $\mathtt{g}^{(1,\frac{1}{2})}_{\ell}$ and $\mathtt{h}^{(1,\frac{1}{2})}_{\ell}$ are integration constants and $x=r/2M-1$. From the regularity of the scalar field, 
we need to set $\mathtt{g}^{(1,\frac{1}{2})}_{\ell}=\mathtt{h}^{(1,\frac{1}{2})}_{\ell}=0$. 
Therefore, 
the scalar field $\varphi^{(1,\frac{1}{2})}$ has only $\ell=0$ mode and the configuration becomes
\begin{align}
    \varphi^{(1,\frac{1}{2})}(r)
    =
    \mathrm{const}.
\end{align}
As in the case of the order $\mathcal{O}(\zeta)$, the time dependence of the scalar field is not allowed at the order $\mathcal{O}(\varepsilon \zeta)$.

The metric of the order $\mathcal{O}(\varepsilon \zeta)$ 
can be obtained by solving the field equations~\eqref{fieldeqesgb}: 
\begin{align}
    g_{t\phi}^{(1,1)}(r)
    = 
    \frac{3}{5}\varepsilon \zeta \frac{M^4 \sin^2 \theta}{r^3}
    \left(
    1 + \frac{140M}{9r} + \frac{10M^2}{r^2} + \frac{16M^3}{r^3} - \frac{400M^4}{9r^4} 
    \right).
\end{align}
This result is same from in~\cite{Pani:2011gy}. 
Thus, the circularity condition holds at this order.

\subsection{$\mathcal{O}(\varepsilon^2 \zeta)$ corrections}

We can show that the scalar field is time independent:  $\varphi^{(2,\frac{1}{2})} = \varphi^{(2,\frac{1}{2})}(r)$ by the same way as the case of the order $\mathcal{O}(\varepsilon \zeta)$. 
The configuration of the scalar field $\varphi^{(2,\frac{1}{2})}$ is given by~\cite{Pani:2011gy,Ayzenberg:2014aka}
\begin{align}
    \varphi^{(2,\frac{1}{2})}(r)
    =
    -  \frac{\varepsilon^2 \alpha}{2Mr}
    \left[
    1 + \frac{M}{r} + \frac{4M^2}{5r^2} + \frac{2M^3}{5r^3} + 
    \frac{28M^2 \cos^2 \theta}{5r^2}
    \left(
    1 + \frac{3M}{r} + \frac{48M^2}{7r^2}
    \right)
    \right].
\end{align}
We choose the homogeneous integration constants so that the scalar field to be regular everywhere. 
The metric of the order $\mathcal{O}(\varepsilon^2 \zeta)$ is given by the same form obtained in~\cite{Ayzenberg:2014aka}.\footnote{
The functions $H_{0\ell}^{(2,1)}$, $H_{2\ell}^{(2,1)}$, and $K_{\ell}^{(2,1)}$
correspond to the functions $H_{0\ell 0}, H_{2\ell 0}$ and $K_{\ell 0}$ in Eqs.(A15)-(A21) in~\cite{Ayzenberg:2014aka}, respectively 
(see also the erratum in~\cite{Ayzenberg:2014aka} because there are some typos).} 
Thus, the spacetime is circular at this order.

\section{summary and discussion}
\label{sec:conclusion}

We have studied slowly rotating black hole solutions up to linear order in coupling constant and quadratic order in the spin in dynamical Chern--Simons gravity and shift symmetric Einstein scalar Gauss--Bonnet gravity. 
In our analysis, we did not assume the circularity condition for the spacetime and the time independence of the scalar fields.
We have shown that in both gravity theories, up to this order, such regular non-circular black hole solutions do not exist at least around the Schwarzschild spacetime. 
We note that the linear time dependence of the scalar fields leads non-circular spacetimes in dynamical Chern--Simons gravity, while the linear time dependence makes the spacetime singular. 
We also have shown that in dynamical Chern--Simons gravity, the time dependence of the scalar field is limited to the linear time dependence from the stationarity of the effective stress energy tensors. 
On the other hand, we have shown in shift symmetric Einstein scalar Gauss--Bonnet gravity, any time dependence of the scalar field is prohibited. 
This is the consequence of the conditions~\eqref{eq:stationaryconesgb} and~\eqref{eq:Tti} which mean that the effective stress energy tensor is stationary and the $ti$-component of the effective stress energy tensor should vanish for $\ell = 0$ mode. 
Our results suggest the non-existence of rotating non-circular black holes in these gravity theories.

In our analysis, 
we first determine the time dependence of the scalar fields from the stationarity of the metric functions and the effective stress energy tensors, and next we solve the field equations.
In both steps, we adapt the slow rotation approximation, and solve the equations at each order of the spin parameter.
Since our analysis method is widely usable, 
it is interesting to use it to find non-circular black hole solutions in other string-inspired gravity theories,
but we leave this to future work.

\section*{ACKNOWLEDGMENTS}
We would like to thank K. Aoki, V. Cardoso, T. Harada, T. Ikeda, W. Ishizaki,  T. Kobayashi, Y. Koga, H. Motohashi, A. Naruko, K. Ogasawara, K. Takahashi, T. Tanaka, and K. Yamada for useful comments and discussions. 
K.N. also thanks the Yukawa Institute for Theoretical Physics at Kyoto University, where this work was developed, for their hospitality. 
This work was supported by the Rikkyo University Special Fund for Research (K.N.) and a MEXT Grant-in-Aid for Scientific Research on Innovative Areas 20H04746 (M.K.).

\appendix

\section{circularity condition}
\label{app:circularity}

In this section, we briefly review the circularity condition~\cite{heusler_1996,wald1984general}. 
The stationary and axisymmetric spacetime possesses an asymptotically timelike Killing vector field $k$ and a spacelike Killing vector field $m$ whose orbit is closed.
We assume that these two Killing vector fields commute.
This implies that we can choose a coordinate system $\{ t, x^1, x^2, \phi \}$ where $k = \partial _t$ and $m = \partial _\phi$ are coordinate basis vectors.
Consequently, the metric components in this coordinate system are independent of $t$ and $\phi$:
\begin{align}
    \mathrm{d} s^2 = g_{\mu \nu} (x^1, x^2) \mathrm{d} x^\mu \mathrm{d} x^\nu.
\end{align}
The spacetime is said to be circular if the two-dimensional surface orthogonal to $k$ and $m$ is integrable.
From the Frobenius's theorem, the integrability condition is given by
\begin{align}
    dk \wedge k \wedge m = dm \wedge m \wedge k = 0.
    \label{eq:appA2}
\end{align}
When these conditions are satisfied, the metric can be decomposed as follows
\begin{align}
    \mathrm{d} s^2 = \sum_{A,B} g_{AB} \mathrm{d} x^A \mathrm{d} x^B + \sum_{a,b} g_{ab} \mathrm{d} x^a \mathrm{d} x^b,
\end{align}
where $A,B$ run $t, \phi$ while $a,b$ run $x^1, x^2$, respectively.
That is, when the spacetime is circular, the $t\, x^a$- and $\phi \, x^a$-components of the metric vanish.
For the stationary, axisymmetric, and asymptotically flat spacetime, 
the conditions~\eqref{eq:appA2}
are equivalent to the Ricci-circular conditions 
\begin{align}
   R(k) \wedge k \wedge m = R(m) \wedge m \wedge k = 0,
\end{align}
where $R(k) = k^{\mu} R_{\mu \nu}$ and $R(m) = m^{\mu} R_{\mu \nu}$. 
In the vacuum solution of general relativity in which $R_{\mu \nu}=0$, the spacetime is always circular.

\section{linear stationary metric perturbation in the Regge--Wheeler gauge}
\label{appB}

In this section, we review the linear stationary metric perturbation around the spherically symmetric spacetime in the Regge--Wheeler gauge briefly. 
The background metric is given by 
\begin{align}
    \mathrm{d} s^2_{\mathrm{BG}}
    = 
    -f(r) \mathrm{d}t^2 
    + \frac{\mathrm{d}r^2}{f(r)} 
    + r^2 \sum_{a,b} \gamma_{ab} \mathrm{d}x^a \mathrm{d}x^b,
\end{align}
where $a, b$ run $\theta, \phi$, and $\gamma_{ab}$ is the metric on $S^2$ 
\begin{align}
    \sum_{a,b} \gamma_{ab} \mathrm{d}x^a \mathrm{d}x^b
    =
    \mathrm{d}\theta^2 + \sin^2 \theta \mathrm{d}\phi^2. 
\end{align}
In the following subsection, we show that the even and odd metric perturbations in the Regge--Wheeler gauge are given by
\begin{align}
    h_{\mu \nu}^{(+)} \mathrm{d}x^{\mu} \mathrm{d}x^{\nu} &= 
    \left( H_0(r) \mathrm{d}t^2 + 2 H_1(r) \mathrm{d}t \mathrm{d}r + H_2(r) \mathrm{d}r^2 \right) \mathbb{S} + 
    2 r^2 K(r) \,\mathbb{S} \sum_{a,b} \gamma _{ab} \mathrm{d}x^{a} \mathrm{d}x^{b},
    \\
    h_{\mu \nu}^{(-)} \mathrm{d}x^{\mu} \mathrm{d}x^{\nu} &= 
    (2  h_{0}(r) \mathrm{d}t + 2  h_{1}(r) \mathrm{d}r) \sum_{a} \mathbb{V}_{a} \mathrm{d}x^{a}.
    \label{eq:oddparityRW}
\end{align}
In the right hand side of the above equations, we should take summation of $\ell$, but we omit to write it
in this paper.
The scalar harmonics $\mathbb{S}$ is defined by regular solutions of the equation 
\begin{align}
    \Big( \sum_{a} \hat D^{a} \hat D_{a} + \ell(\ell+1) \Big) \mathbb{S}=0, 
\end{align}
where $\hat D_{a}$ is the covariant derivative on $S^2$ and $\ell = 0,1,2,3,\cdots$. 
The scalar harmonics $\mathbb{S}$ can be written by $\mathbb{S} = \sum_m c_m Y_{\ell m}$, where $Y_{\ell m}$ is the spherical harmonics and $c_m$ are constants. If we impose the spacetime symmetry along $\partial_\phi$, we can set $\mathbb{S} = Y_{\ell 0}$.
The vector harmonics $\mathbb{V}_{a}$ is expressed as $\mathbb{V}_{a} = \sum_{b} \hat \epsilon_{a}{}^{b} \hat D_{b} \mathbb{S}$ where $\hat \epsilon_{ab}$ is the Levi--Civita tensor on $S^2$. 
The vector harmonics $\mathbb{V}_{a}$ satisfies
\begin{align}
    \Big( \sum_{b} \hat D^{b} \hat D_{b} + \ell (\ell +1)-1 \Big) \mathbb{V}_a = 0,
\end{align}
with $\sum_{a} \hat D^{a} \mathbb{V}_{a}=0$. 
Because $\mathbb{V}_a=0$ for $\ell=0$, 
we only need to consider $\ell \ge 1$ for the odd parity metric perturbation~\eqref{eq:oddparityRW}. 
We note that $h_{1}|_{\ell =1} = H_{1}|_{\ell=1} = K|_{\ell =0} = K|_{\ell=1}=0$.

\subsection{even parity stationary perturbation}

The general even parity stationary metric perturbation can be written as 
\begin{align}
    h^{(+)}_{\mu \nu} &\mathrm{d}x^{\mu} \mathrm{d}x^{\nu} 
    = 
    \left( H_0(r) \mathrm{d}t^2 + 2 H_1(r) \mathrm{d}t \mathrm{d}r + H_2(r) \mathrm{d}r^2 \right) \mathbb{S} \,+
    \notag\\
    & 
    \left( 2 r H_{t \Omega}\, \mathrm{d}t + 2 r H_{r \Omega} \, \mathrm{d}r \right) \sum_{a} \mathbb{S}_{a} \mathrm{d} x^{a}  +
    2r^2 \sum_{a,b} 
    \left( K(r) \mathbb{S}\, \gamma _{ab} + 
    K_2(r) \mathbb{S}_{ab} 
    \right)\mathrm{d}x^{a} \mathrm{d}x^{b}.
\end{align}
The two quantities $\mathbb{S}_{a} (\ell \ge 1)$ and $\mathbb{S}_{ab} (\ell \ge 2)$ are expressed by 
\begin{align}
    \mathbb{S}_{a} = - \frac{1}{\sqrt{\ell(\ell +1)}} \hat D_{a} \mathbb{S}, \,\,\,\,
    \mathbb{S}_{ab} = \frac{1}{\ell(\ell +1)} \hat D_{a} \hat D_{b}\, \mathbb{S} + \frac{1}{2} \gamma_{ab}\, \mathbb{S}. 
\end{align}
Note that $\mathbb{S}_a = 0$ for $\ell=0$ and $\mathbb{S}_{ab}=0$ for $\ell=0,1$. 

We perform an infinitesimal coordinate transformation $x^{\mu} \to x^{\mu} + \xi ^{\mu}$. 
The gauge functions $\xi_{\mu}$ for even stationary metric perturbation are given by 
\begin{align}
    \xi_{t} = T(r) \mathbb{S}, \,\,\,\,
    \xi_{r} = R(r) \mathbb{S}, \,\,\,\,
    \xi_{a} = \xi_{\Omega} (r) \mathbb{S}_{a}. 
\end{align}
The transformation laws are given by 
\begin{align}
    H_0(r) &\to H_0(r) - f(r)f'(r) R(r), 
    \\
    H_1(r) &\to H_1(r) + T'(r) - \frac{f'(r)}{f(r)} T(t), 
    \\
    H_{2}(r) &\to H_{2}(r)  + 2 R'(r) + \frac{f'(r)}{f(r)} R(r),  
    \\
    H_{t\Omega}(r) &\to H_{t\Omega}(r) - \frac{\sqrt{\ell(\ell + 1)}}{r} T(r), 
    \\
    H_{r\Omega}(r) &\to H_{r\Omega}(r) - \frac{\sqrt{\ell(\ell + 1)}}{r^2} R(r) -\frac{2}{r^2} \xi_{\Omega}(r) + \frac{\xi_{\Omega}'(r)}{r}, 
    \\
    K(r) &\to K(r) + \frac{f(r)}{r} R(r) + \frac{\sqrt{\ell(\ell +1)}}{2r^2} \xi_{\Omega}(r), 
    \\
    K_2(r) &\to K_2(r) - \frac{\sqrt{\ell(\ell +1)}}{r^2} \xi_{\Omega}(r), 
    \end{align}
where prime denotes the derivative with respect to $r$. 
Note that 
We can see that for $\ell \ge 2$, $K_2$, $H_{t\Omega}$, and $H_{r\Omega}$ can be set to zero by solving algebraic equations for $\xi_{\Omega}$, $R$, and $T$. For $\ell =1$, we can set $K$, $H_{t\Omega}$, and $H_{r\Omega}$ to be zero. 
In the case of $\ell = 0$, where the relevant gauge functions are $T$ and $R$, 
we can use these gauge functions to set $K=H_{1}=0$.

\subsection{odd parity stationary perturbation}

The general odd parity stationary metric perturbation can be written as 
\begin{align}
    h^{(-)}_{\mu \nu} \mathrm{d}x^{\mu} \mathrm{d}x^{\nu} = 
    \left( 2h_{0}(r)\, \mathrm{d}t + 2 h_{1}(r)\, \mathrm{d}r \right) 
      \sum_{a} \mathbb{V}_{a} \mathrm{d} x^{a}  
    + 2\, r^2\, h_{\Omega}(r) \sum_{a,b} \mathbb{V}_{ab} \mathrm{d} x^{a} \mathrm{d} x^{b} . 
    \label{eq:oddmetric}
\end{align}
The symmetric trace-free tensor $\mathbb{V}_{ab}$ is expressed as 
\begin{align}
    \mathbb{V}_{ab} = - \frac{\hat D_{(a} \mathbb{V}_{b)}}{\sqrt{\ell(\ell + 1)-1}}.
\end{align}
We note that $\mathbb{V}_{ab}=0$ for $\ell = 1$.

For the odd parity stationary metric perturbation, the gauge functions $\xi_{\mu}$ can be written as 
\begin{align}
    \xi_{t} = \xi_{r} = 0,\,\,\,
    \xi_{a} = \xi_{\Omega}(r) \mathbb{V}_a, 
\end{align}
where $\xi_{\Omega}$ is an arbitrary function of $r$. 
The transformation laws are given by
\begin{align}
    h_{0}(r) &\to h_{0}(r), 
    \label{eq:h0trans} 
    \\
    h_{1}(r) &\to h_{1}(r) +
    \xi_{\Omega}'(r) - \frac{2}{r} \xi_{\Omega}(r), 
    \label{eq:h1trans}
    \\
    h_{\Omega}(r) &\to h_{\Omega}(r) - \frac{\sqrt{\ell(\ell+1)-1}}{r^2} \xi_{\Omega}(r).
    \label{eq:homegatrans}
    \end{align}
From these transformation laws, 
we can set $h_{\Omega}=0$ for $\ell \ge 2$ modes. 
For $\ell =1$, 
because $\mathbb{V}_{ab}=0$, $h_{\Omega}$ does not appear in the odd parity metric perturbation~\eqref{eq:oddmetric}. 
For this reason, we need to consider only Eqs.~\eqref{eq:h0trans} and~\eqref{eq:h1trans}, 
and we set $h_1=0$ for $\ell = 1$ by choosing $\xi_{\Omega}$ appropriately.

\section{homogeneous solution of stationary metric perturbation}
\label{appC}

In this section, we discuss the homogeneous solution of the stationary metric perturbation, i.e., 
we solve $G_{\mu \nu}=0$. 
We express the metric perturbation in the Regge--Wheeler gauge as shown in~\eqref{eq:perturbation}.

\subsection{even parity perturbation}

We start with the discussion on the even parity metric perturbations $H_{0\ell}$, $H_{1\ell}$, $H_{2\ell}$, and $K_{\ell}$. 
For $\ell = 0$, 
we only need to consider $H_{0 0}$ and $H_{2 0}$. 
We solve $G_{tt}=0$ with respect to $H_{20}$ and we use it to solve $G_{rr}=0$ with respect to $H_{00}$. 
The solutions are given by 
\begin{align}
    H_{00}(r) 
    &= 
    \frac{\mathtt{a}}{rf} + \mathtt{b}, \\
    H_{20}(r)
    &= 
    \frac{\mathtt{a}}{rf},
\end{align}
where $\mathtt{a}$ and $\mathtt{b}$ are integration constants and $f=1-2M/r$. 
By imposing that the spacetime is asymptotically flat and the mass of the black hole measured at the spatial infinity is $M$, 
we can set $\mathtt{a}=\mathtt{b}=0$.

For $\ell=1$, 
the non-vanishing components of the metric perturbation are $H_{01}$, $H_{11}$, and $H_{21}$. 
From $G_{rt}=0$, we can see $H_{11}=0$. By the same procedure in the case of $\ell=0$, 
we obtain $H_{01}$ and $H_{21}$ as
\begin{align}
    H_{01}(r) 
    &= 
    \frac{\mathtt{c}}{3r^2 f^2} + \mathtt{d}\, rf,
    \\
    H_{21}(r)
    &= 
    \frac{\mathtt{c}}{r^2 f^2},
\end{align}
where $\mathtt{c}$ and $\mathtt{d}$ are integration constants. 
In order to satisfy $G_{r\theta}=0$, 
we need to set $\mathtt{d}=0$. 
We also need to $\mathtt{c}=0$ so that the spacetime is regular at $r=2M$. 

For $\ell \ge 2$, we can see $H_{1\ell}=0$ from $G_{rt}=0$. 
Combining $G_{\theta \theta}=0$ and $G_{\phi \phi}=0$, we consider $G_{\theta\theta}+G_{\phi \phi}/\sin^2\theta=0$. 
The general solution of this equation is 
\begin{align}
    H_{0 \ell}(r) = H_{2 \ell}(r).
\end{align}
Then, we need to solve the coupled differential equations with respect to $H_{2 \ell}$ and $K_{\ell}$. 
The general solutions are given by
\begin{align}
    H_{2\ell}(x)
    &= 
    \mathtt{e} P^{2}_{\ell}(2x+1) + 
    \mathtt{f} Q^{2}_{\ell}(2x+1), 
    \\
    K_{\ell}(x)
    &=
    \frac{\mathtt{e}}{2(\ell+2)x(x+1)}
    \left[
    \{2x(x(\ell+2)+\ell+1)-1\}P^{2}_{\ell}(2x+1)
    + P^{2}_{\ell+1}(2x+1)
    \right]
    \notag\\
    &+
    \frac{\mathtt{f}}{2(\ell+2)x(x+1)}
    \left[
    \{2x(x(\ell+2)+\ell+1)-1\}Q^{2}_{\ell}(2x+1)
    + Q^{2}_{\ell+1}(2x+1)
    \right],
\end{align}
where $\mathtt{e}$, $\mathtt{f}$ are integration constants, $P^{m}_{n}(x)$ and $Q^{m}_{n}(x)$ are the associated Legendre polynomial of the first kind and second kind, respectively, and $x=r/2M-1$. 
When we require that the spacetime is regular everywhere, we need to set $\mathtt{e}=\mathtt{f}=0$.

\subsection{odd parity perturbation}
We discuss the odd parity metric perturbation $h_{0 \ell}$ and $h_{1 \ell}$. 
It is sufficient to consider the case of $\ell \ge 1$ for odd parity metric perturbation. From $G_{r\phi}=0$, we have $h_{1 \ell}=0$ for $\ell \ge 1$. 

For $\ell =1$, 
the expression for $h_{01}$ can be obtained from $G_{t\phi}=0$ by
\begin{align}
    h_{01}(r) 
    = 
    \frac{\mathtt{g}}{r} + \mathtt{h}\,r^2,
\end{align}
where $\mathtt{g}$ and $\mathtt{h}$ are integration constants. From the asymptotic flatness of the spacetime, we need to set $\mathtt{h}=0$. 
When we deal with the perturbation around the Schwarzschild black hole, the constant $\mathtt{g}$ is interpreted as the amplitude of the angular momentum of the black hole. 
When we deal with the perturbation around the Kerr black hole, we set $\mathtt{g}=0$ because we consider only the case where the amplitude of the angular momentum of the black hole does not change.  

For $\ell \ge 2$, 
the general solution of $h_{0\ell}$ can be obtained by~\cite{Pani:2011vy}
\begin{align}
    h_{0 \ell} (r) 
    = 
    \mathtt{i} \frac{r^2}{4M^2} 
    \,_{2}F_{1}(1-\ell, \ell +2, 4; \frac{r}{2M})
    + 
    \mathtt{j}\,
    G^{20}_{20} 
    \left( 
    \frac{r}{2M}
     \left|
    \begin{array}{c}
      1-\ell, \ell+2 \\
      -1, 2
    \end{array}
    \right.
    \right),
    \label{eq:meijerg}
\end{align}
where $\mathtt{i}$ and $\mathtt{j}$ are integration constants while $_{2}F_{1}$ and $G$ are the hypergeometric function and the the Meijer function. 
The first term is regular at $r=2M$ but diverge at the infinity while the second term is regular at the infinity but diverge at $r=2M$. 
For the spacetime to be regular everywhere, we need to set $\mathtt{i}=\mathtt{j}=0$.


\bibliographystyle{JHEP}
\bibliography{bibliography}

\end{document}